\documentclass[a4paper,fleqn,usenatbib]{mnras}

\usepackage{mathptmx}

\usepackage[T1]{fontenc}
\usepackage{ae,aecompl}

\pdfoutput=1

\usepackage{graphicx}	
\usepackage{amsmath}	
\usepackage{amssymb}	

\newcommand\ca{\citeauthor}
\newcommand\cy{\citeyear}
\newcommand\vbr{v_{{1}_{R}}}
\newcommand\vbp{v_{{1}_{\phi}}}
\newcommand\vbz{v_{{1}_{z}}}
\newcommand\bbr{B_{{1}_{R}}}
\newcommand\bbp{B_{{1}_{\phi}}}
\newcommand\bbz{B_{{1}_{z}}}
\newcommand\bsr{B_{{0}_{R}}}
\newcommand\bsz{B_{{0}_{z}}}

\title[MRI in Misaligned Protostellar Discs]
{Magnetorotational Instability in Diamagnetic, Misaligned Protostellar Discs}

\author[E. Devlen, A. Ulubay and E. R. Pek\"{u}nl\"{u}]
{Ebru Devlen$^{1}$\thanks{E-mail: devlen@gmail.com, aulubays@istanbul.edu.tr, pekunlu@gmail.com} 
Ayse Ulubay$^{2,3}$
and E. Rennan Pek\"{u}nl\"{u}$^{1}$
\\
$^{1}$Faculty of Science, Department of Astronomy and Space Sciences, Ege University, 35100, Bornova/Izmir, Turkey\\
$^{2}$Faculty of Science, Department of Physics, Istanbul University, 34134, Vezneciler, Istanbul, Turkey\\
$^{3}$Feza G\"{u}rsey Center for Physics and Mathematics, Bo\u{g}azi\c{c}i University, 34684, \c{C}engelk\"{o}y, Istanbul, Turkey}

\date{Accepted 2019 November 24. Received 2019 October 25; in original form 2018 December 18}

\pubyear{2019}

\begin{document}
\maketitle

\label{firstpage}
\begin{abstract}
 In the present study, 
we addressed the question of how the growth rate of the magnetorotational instability is modified 
when the radial component of the stellar dipole magnetic field is taken into account in addition 
to the vertical component. Considering a fiducial radius 
in the disc where diamagnetic currents are pronounced, we carried out a linear stability analysis to obtain  
the growth rates of the magnetorotational instability for various parameters such as the ratio of the 
radial-to-vertical component and the gradient of the magnetic field, the Alfvenic Mach number and the 
diamagnetization parameter. Our results show that the interaction between the diamagnetic 
current and the radial component of the magnetic field increases the growth rate of the 
magnetorotational instability and generates a force perpendicular to the disc plane which may induce a torque. 
It is also shown that considering the radial component of the magnetic field and taking into account a
radial gradient in the vertical component of the magnetic field causes 
an increase in the magnitudes of the growth rates of both the axisymmetric ($m=0$) and 
the non-axisymmetric ($m=1$) modes. 

\end{abstract}
\section{Introduction}
\label{sec:intro} 

The transport of angular momentum through accretion discs is a problem yet to be solved. 
Among several mechanisms proposed, the magnetorotational instability (MRI), which 
relies on the existence of a weak magnetic field in a disc exhibiting a decreasing 
rotation profile, seems to be the most promising one (\ca{Balbus91} \cy{Balbus91}; BH91 hereafter).
When protoplanetary discs are considered, the migration of protoplanetary cores
which then form the planets, depends on basic disc parameters such as the surface density
and temperature. Both of these quantities altered are in the presence of MRI \citep{Kretke10}.

The typical fastest growing waves for poloidal fields have growth rates of
about $0.75 \Omega$, where $\Omega$ is the local Keplerian angular velocity (BH91). 
This value, however, may be somewhat altered when different magnetic field configurations
are considered. When the magnetic field has a radial component, a toroidal field is generated
and as long as this toroidal field is small, the MRI is not affected \citep{Balbus01}.
\cite{Terquem96} studied the stability of discs with purely toroidal magnetic 
fields analytically and numerically. They found that such discs are always unstable, 
and the typical growth rates of the instability for stronger fields are of the order of the
Keplerian rotation frequency of the disc. \cite{Pessah05} carried out a linear
stability analysis for the MRI relaxing the condition of weak magnetic fields.
When radially constant strong toroidal fields are taken into account MRI is stabilized
and a new family of instabilities arise. When the magnetic field is assumed to have only a vertical
component but with an azimuthal dependence, the growth rate of the MRI increases \citep{Dogan17}.

\ca{Devlen07} (\cy{Devlen07}; DP07 hereafter) carried out a linear analysis of the MRI in the presence of the 
diamagnetic effect, taking into account the Hall term for protostellar discs. Their calculations show that the 
diamagnetic effect extends the range of unstable modes to shorter wavelengths and also causes the maximum 
growth rate of the instability exceed the local Oort-A value. 

In the present work, we follow the analysis of DP07 and carry out a linear stability analysis for the MRI 
for diagmagnetic protostellar discs taking into account the radial component of the magnetic 
field in addition to the vertical one. In Section \ref{sec:prelim} we introduce the basic equations we use in our 
analysis. In  Section \ref{sec:linearized} we provide the linearized magnetohydrodynamic (MHD) equations obtained 
from our calculations. The stability criterion and the maximum growth rates for our model parameters are given in Section \ref{sec:results}. 
In Section \ref{sec:discussion} we present a discussion and in Section \ref{sec:summary} we summarize our results.

\section{Preliminaries and Basic Equations}
\label{sec:prelim} 

\begin{figure}
\centering
\includegraphics[width=8cm]{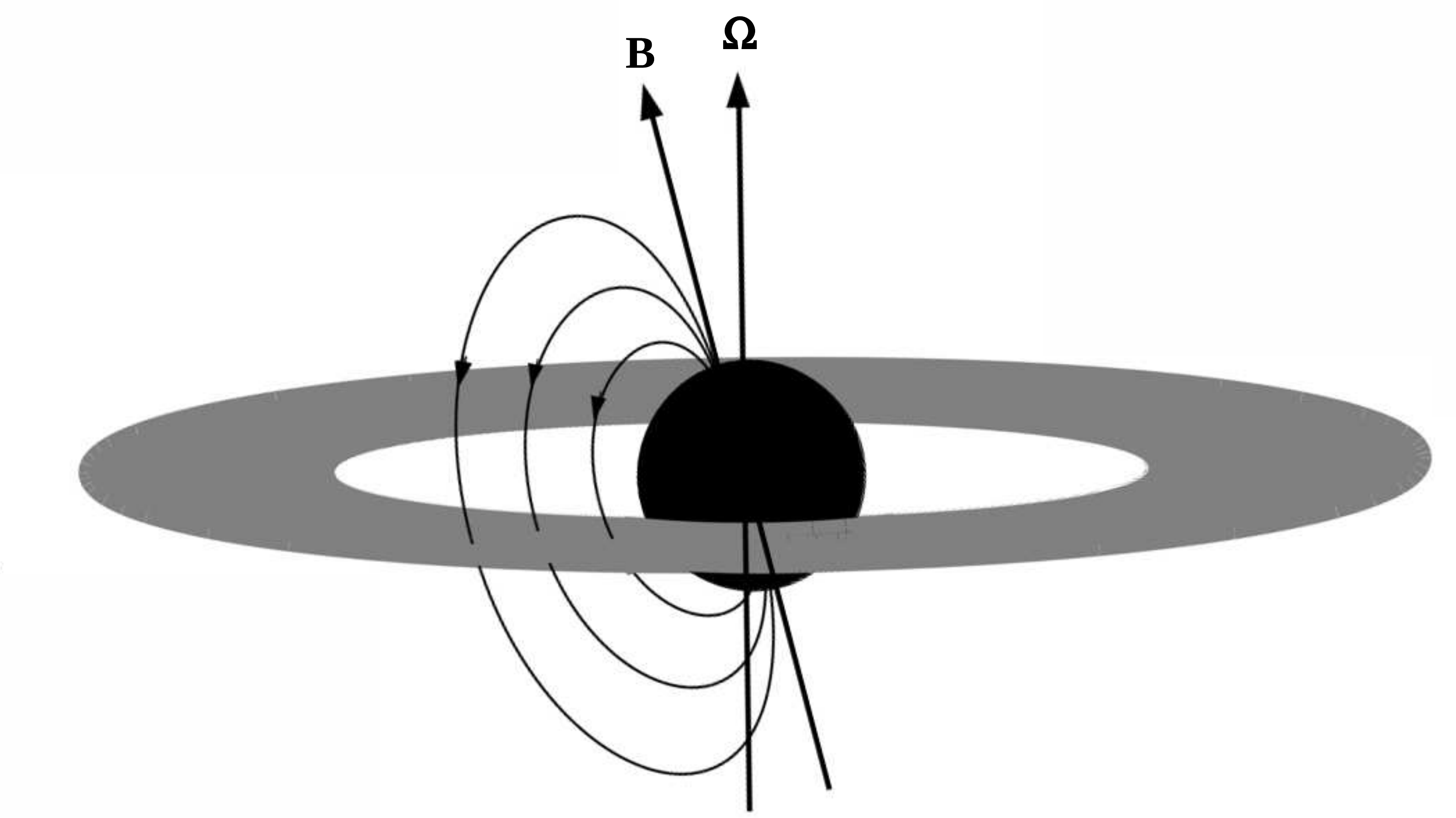}
\caption[Figure01]
{Sketch of the model. A magnetic star is surrounded by a geometrically thin disc 
  threaded by a poloidal magnetic field. The rotation axis of the disc and the magnetic
  field axis are misaligned.}
\label{fig:modeldisc}
\end{figure}

We work in standard cylindrical coordinates (R, $\phi$, z) assuming that a diamagnetic disc rotates around a magnetic 
star with angular velocity 
$\Omega$  which is tilted with respect to the magnetic moment axis. A 
sketch of our model is shown in Fig. \ref{fig:modeldisc}.

When the magnetic field has a radial component $B_{R}$, shear forces in the disc fluid produce a toroidal
field component which grows linearly with time \citep{Balbus91, Blaes94}.  We therefore consider an
equilibrium magnetic field ${{\bf B}}_{0}$ with only radial and vertical components such
that ${{\bf B}}_{0} = (B_{R}(R), 0, B_z({R})) = (\kappa B_{z}(R), 0, B_z({R}))$. In the remainder of this
paper we will refer to these components simply as $\bsr$ and $\bsz$, bearing in mind their dependence on $R$.
As in DP07, we assume that the diamagnetic current is produced by the equatorially trapped plasma particles in 
closed field lines. Neither laminar magnetic torques nor the gravitational ones are relevant in 
the present study. Readers may refer to \cite{Armitage11} for the papers dealing with gravitational torques.

We start by writing the basic MHD equations for a differentially rotating magnetized accretion disc fluid. Given below 
are the equations of continuity,
momentum conservation, magnetic induction and current density, respectively 

\begin{equation}
\frac{\partial \rho}{\partial t} + \nabla \cdot (\rho {\bf{v}}) = 0,
\end{equation}

\begin{equation}
\rho \frac{\partial {\bf{v}} } {\partial t} + (\rho {\bf{v}} \cdot \nabla) {\bf{v}} = -\nabla P + \frac{1}{c} {\bf{J}} \times {\bf{B}},
\end{equation}

\begin{equation}
\frac{\partial {\bf{B}} }{\partial t} =\nabla \times ({\bf{v}} \times {\bf{B}}),
\end{equation}

\begin{equation}
  {\bf {J}} = {\bf {J_{\rm ext}}} +{\bf {J_{\rm mag}}} = \frac {c}{4 \pi} \nabla \times {\bf {B}} + c \nabla \times {\bf {M}},
\end{equation}
where $\rho$ is the mass density, ${\bf{v}} =(0, v_{\phi},0) = (0, R \Omega(R),0)$ is the fluid velocity, $P$ is 
the pressure, {{\bf B}} is the magnetic field, {{\bf J}} is the
electrical current density, c is the speed of light in vacuum and ${\bf{M}}$ is the magnetization, the details of 
which are given in the next subsection.

\subsection{Diamagnetic effect}
\label{sec:diamagnetic}
As is well known, charged particles in a magnetic field acquire three periodic motions: 
$i$) circular motion around the magnetic field lines; 
$ii$) bounce motion between magnetic mirrors and 
$iii$) drift motion around the central object (protostar) \citep{Schulz74}.

Any free charged particle (in our case, diamagnetic current producing electrons) may enter the dipole 
magnetic field with a velocity vector {\bf{v}} and generally shows the three periodic motions mentioned above. 
But, there are special cases; for instance, if an electron enters the magnetic field perpendicularly  at the 
magnetic equator, it is trapped at the magnetic equator and acquires only the first and the third periodic 
motions, i.e., circular motion around the magnetic field lines and drift around the central object. Those 
electrons are called ``equatorially trapped particles''. 

When the electrons move in a global external magnetic field, {\bf{H}}, they create currents which in 
turn generate a local magnetic field {\bf{M}} with a direction opposite to that of {\bf{H}}. This local 
magnetic field, or magnetization, can be written as \citep{Singal86, Bodo92}
\begin{equation}
{\bf{M}}=-\frac{2 {\bf{B}} }{3B^2}W_k,
\label{eq:magnetization}
\end{equation}
where $W_{k}=nmv_{\perp}/2$ is the kinetic energy density of non-relativistic 
electrons with number density $n$, mass $m$ and perpendicular velocity $v_{\perp}$. The net 
magnetic field in the disc is the sum of the magnetization and the external field, and is written as \citep{Singal86}
\begin{equation}
    {\bf{B}}={\bf{H}}+4 \pi {\bf{M}} = {\bf{H}}-\frac{8 \pi}{3}W_k \frac{{\bf{B}}}{B^2} ={\bf{H}}-\frac{1}{3} \frac{W_k}{W_B}{\bf{B}}.
    \label{eq:totalB}
\end{equation}
Here, $W_B=B^2/ 8 \pi$ is the magnetic energy density.  At a fiducial radius the strength of the diamagnetic current will depend on the magnetization, $\epsilon$, defined as  $\epsilon=W_k/3W_B$ as in DP07. There, it was found that the maximum value of $\epsilon$ is 0.5 by considering a solution to the quadratic equation obtained by 
taking the scalar product of equation \ref{eq:totalB} with {\bf{B}} and allowing for the solution that 
allows magnetization (see also \cite{Singal86, Bodo92}). We adopt this limiting value 
of $\epsilon$ in our analysis of growing MRI modes.

We assume that in the protostellar disc electron fluid is frozen in protostellar magnetic field lines. This means that electrons and magnetic field lines co-rotate. Since electrons are trapped in the equatorial part of the magnetic field lines and exhibit both gyration around the field lines and drift in the azimuthal direction, they generate a magnetic field. This diamagnetism may be global or local in the disc (DP07). However, electrons can be frozen to the magnetic field only in the inner, hot regions, therefore our study focuses on those parts of the disc.

\section{Linearized MHD Equations}
\label{sec:linearized}
Let $\rho_1$, ${{\bf v}}_1$, ${{\bf B}}_1$ and $P_1$ denote small perturbations
to the equilibrium quantities $\rho_0$, ${{\bf v}}_0$, ${{\bf B}}_0$ and $P_0$, respectively. 
We adopt the Boussinesq approximation and consider 
perturbations of the form ${\rm {exp}}({i{\bf{k}}\cdot{\bf{r}}- i \omega t})$,
where ${\bf{k}}=(0,m/R,k_{z})$ is the wave vector and 
$\omega = \operatorname{Re} (\omega)+ \it{i} \operatorname{Im}(\omega)$
is the frequency of a wave mode. The linearized continuity equation then reads

\begin{equation}
\frac{m}{R} {v_{1}}_{\phi} + k_z v_{1_{z}}=0.
\label{eq:cont}
\end{equation}

\noindent
We find the ($R$, $\phi$, $z$) components of the linearized momentum equation as 

\begin{multline}
-i (\omega - m \Omega) \vbr - 2 \Omega \vbp 
+\Big(\frac{1}{4 \pi \rho} (i k_z \bsz)(\epsilon-1) \Big) \bbr \\
+ \Big[\frac{1}{4 \pi \rho} \Big((1-\epsilon) \frac{d\bsz}{dR}-  2 i k_z \epsilon \kappa \bsz \Big) \Big] \bbz =0,
\label{eq:pmomR}
\end{multline}

\begin{multline}
-  i (\omega - m \Omega) \vbp
  + \frac{1}{2} \frac{{\kappa_{e}}^2} {\Omega} \vbr 
  + \frac{i m}{R \rho} P_1 \\
  \hspace{1.89cm}
+\Big[\frac{1}{4 \pi \rho} \Big(
\frac{1}{R} i m \bsz (\kappa(1-\epsilon)) \Big)\Big] \bbr \\
\hspace{1.89cm}
+  \Big[\frac{1}{4 \pi \rho} \Big(
i k_z \bsz (\epsilon-1)-2 \epsilon \kappa \frac{d \bsz}{dR}\Big)\Big] \bbp \\
  +\Big[\frac{1}{4 \pi \rho} \Big(
\frac{1}{R} i m \epsilon \bsz (\epsilon+2 \kappa^2 \epsilon +1)
\Big)\Big] \bbz =0,
\label{eq:pmomphi}
\end{multline}

\begin{multline}
-i (\omega - m \Omega) \vbz +  \Big(\frac{i k_z}{\rho}\Big) P_1 \\
\hspace{2.5cm}
+\Big[ \frac{1}{4 \pi \rho} \Big(
 i k_z \bsz \kappa (1-\epsilon) - \frac{d \bsz}{dR} (1+\epsilon) \Big) \Big]\bbr \\
 +\Big[ \frac{1}{2 \pi \rho} \Big(
\epsilon \kappa \frac{d \bsz}{dR} + i k_z \epsilon \kappa^2 \bsz \Big]\bbz =0, 
\label{eq:pmomz}
\end{multline}

\noindent
 and the ($R$, $\phi$, $z$) components of the linearized induction equation as
\begin{align}
\begin{split}\label{eq:pinductR}
{}& -i (\omega - m \Omega) \bbr - \Big(i k_z \bsz + \kappa \frac{d \bsr}{d R} \Big) \vbr =0,
\end{split}\\
\begin{split}\label{eq:pinductphi}
{}& -i (\omega - m \Omega) \bbp - \Big( \frac{d \Omega}{d ln R} \Big)\bbr
- (i k_z \bsz) \vbp=0,
\end{split}\\
\begin{split}\label{eq:pinductz}
{}& - i (\omega - m \Omega) \bbz 
+ \Big( \frac{d \bsz}{d R} \Big) \vbr
- (i k_z \bsz) \vbz 
=0,
\end{split}
\end{align}
where $ \kappa = {B_{\rm 0_{R}}}/{B_{\rm 0_{z}}}$ and $\kappa_e$ is the epicyclic frequency. 

\section{Dispersion Relations}
\label{sec:results}

In order to obtain the dispersion relations we construct the coefficient matrix $\mathcal C$ from the linearized MHD 
equations \ref{eq:cont} - \ref{eq:pinductz} and solve $\rm det(\mathcal C)=0$ for axisymmetric ($m=0$) and 
non-axisymmetric disturbances. The general dispersion relation is a fifth order polynomial with complex 
coefficients and it reduces to a quartic for the $m=0$ mode. To obtain the solutions to the dispersion relations, we 
make use of the Numerical Algorithms Group's routines C02ANFE (for the $m=0$ case) and C02AFFE (for the non-axisymmetric 
case).

The form of the perturbations we consider imply that when $\operatorname{Im}(\omega) > 0$ the dispersion 
relation admits solutions growing with time, i.e. the disc is unstable. In the following subsections \ref{sec:Axisymetric} 
and \ref{sec:Nonaxisymetric} 
we show these growing solutions for the axisymmetric and the non-axisymmetric cases, respectively.

\subsection{Axisymmetric mode}
\label{sec:Axisymetric}
The dispersion relation for the axisymmetric mode is
\begin{equation}
  {{\sigma}}^4 + a_0 {{\sigma}}^2 +b_0 = 0,
  \label{eq:dispersionm0}
\end{equation}
with
\begin{equation}
a_0 = \frac{G^2}{{M_A}^2}(1-\epsilon) -2X^2(1-\epsilon)-i \frac{\kappa X G}{M_A}(1-\epsilon)-\eta^2,
\end{equation}
and 
\begin{multline}
b_0 = i\frac{2 \kappa G^3 X \epsilon}{{M_A}^3}(1-\epsilon)-\frac{X^2 G^2}{{M_A}^2}(\epsilon-1)^2+
X^4(\epsilon-1)^2 \\
+2X^2 \alpha (1-\epsilon) +i \frac{ \kappa X^3 G}{M_A}(\epsilon-1)^2 + 
i \frac{2 X \kappa G \alpha}{M_A}(1-3 \epsilon) \\
+ \frac {2 X^2 \epsilon G^2 \kappa^2}{{M_A}^2}(\epsilon +1)+ 
\frac{4 \kappa^2 G^2 \epsilon \alpha}{{M_A}^2},
\end{multline}
where
$\sigma = \omega / \Omega$, $ G =  {d {\rm{ln}} B_{0_{z}}}/{d {\rm{ln}} R}$, $ M_A = {v_{\phi}}/{v_A}$ is 
the Alfvenic Mach number, $X = k_z v_A / \Omega$, $\eta = \rm {\kappa_e}/{\Omega}$ and $\alpha = {d {\rm{ln}} \Omega}/{d \ln R}$.
\subsubsection{Stability criterion}
\label{sec:criterion}
We consider a disc with a magnetic field component only in the $z-$direction,
i.e. $\kappa=0$ and $B_R=0$, and study its stability
using the standard Routh-Hurwitz theorem valid for polynomials with real coefficients. The necessary and
sufficient condition for stability is given as $b_0 \ge 0$. This gives
\begin{equation}
 [k_z^2 v_A^2 (1-\epsilon)] \Big[k_z^2 v_A^2 (1-\epsilon)-v_A^2
  \Big( \frac{\partial {\rm{ln}} \bsz}{\partial R} \Big)^2
  (1-\epsilon)
  +\frac{\partial \Omega^2}{\partial {\rm{ln}} R} \Big] \ge 0.
\label{eq:crit}
\end{equation}
The first factor in the above inequality represents the magnetic tension force and hence is always positive.
Therefore the stability condition reduces to the below inequality
\begin{equation}
  k_z^2 v_A^2 (1-\epsilon) \ge
 \Big| \frac{\partial \Omega^2}{\partial {\rm{ln}} R} \Big|
+ v_A^2 \Big( \frac{\partial {\rm{ln}} \bsz}{\partial R} \Big)^2  (1-\epsilon).
\label{eq:crit_fin}
\end{equation}
The second term on the right hand side of the above inequality \ref{eq:crit_fin} results from the diamagnetic current which generates the gradient in the vertical component of the magnetic field in the disc and represents the deviation from the standard BH91 condition. This term puts an 
additional destabilization to the disc compared to standard MRI.

The maximum growth rate derived from equation \ref{eq:dispersionm0} is given by 
\begin{equation}
  |\omega_{\rm max} | =\frac{1}{2}
  \Big| \frac{\partial \Omega}{\partial {\rm{ln}} R} \Big|+
  \frac {1}{4 \Omega}
  \Big( \frac{\partial {\rm{ln}} \bsz}{\partial R} \Big)^2 v_A^2 (1-\epsilon)
\label{eq:maxomega}
\end{equation}
which occurs when
\begin{equation}
  (k_{z} v_A )^2_{\rm max} =  \frac{\Omega^2}{1-\epsilon}
\Big[1 - \frac{1}{4 \Omega^4} \Big( \frac{1}{2} {\kappa_e}^2 - \frac{1}{2} \Big( \frac{\partial {\rm ln} \bsz}{\partial R} \Big)^2 
{v_A}^2 (1-\epsilon) \Big)^2 \Big].
\end{equation}

Inequality \ref{eq:crit_fin} and equation \ref{eq:maxomega} clearly show that the extra agent for destabilization increases the maximum growth rate of instability.

\subsubsection{Growth rates}
\label{sec:growthm0}

\begin{figure*}
\centering
  \begin{minipage}[htbp!]{0.4\textwidth}
    \includegraphics[width=\linewidth]{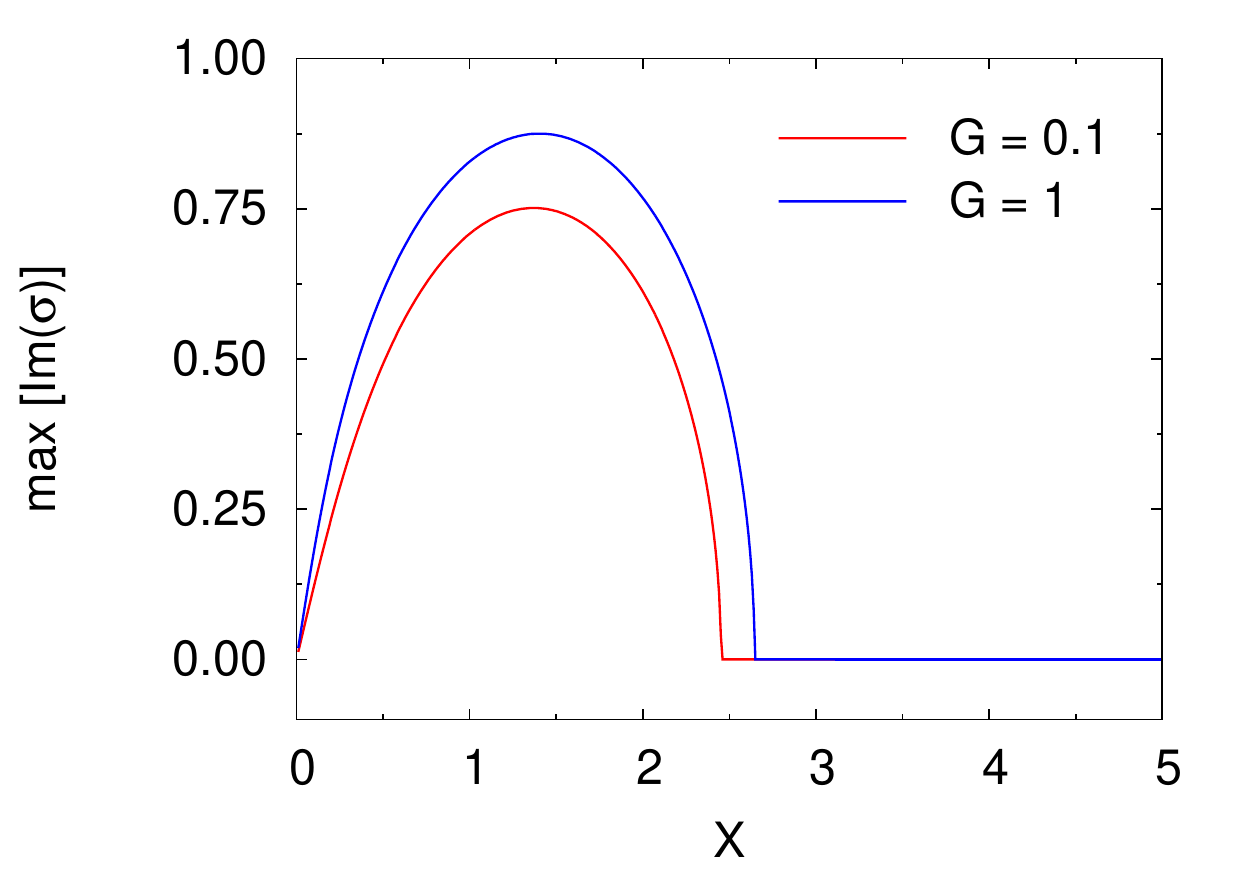}
  \end{minipage}
  \hspace{0.7cm}
  \begin{minipage}[htbp!]{0.4\textwidth}
    \includegraphics[width=\linewidth]{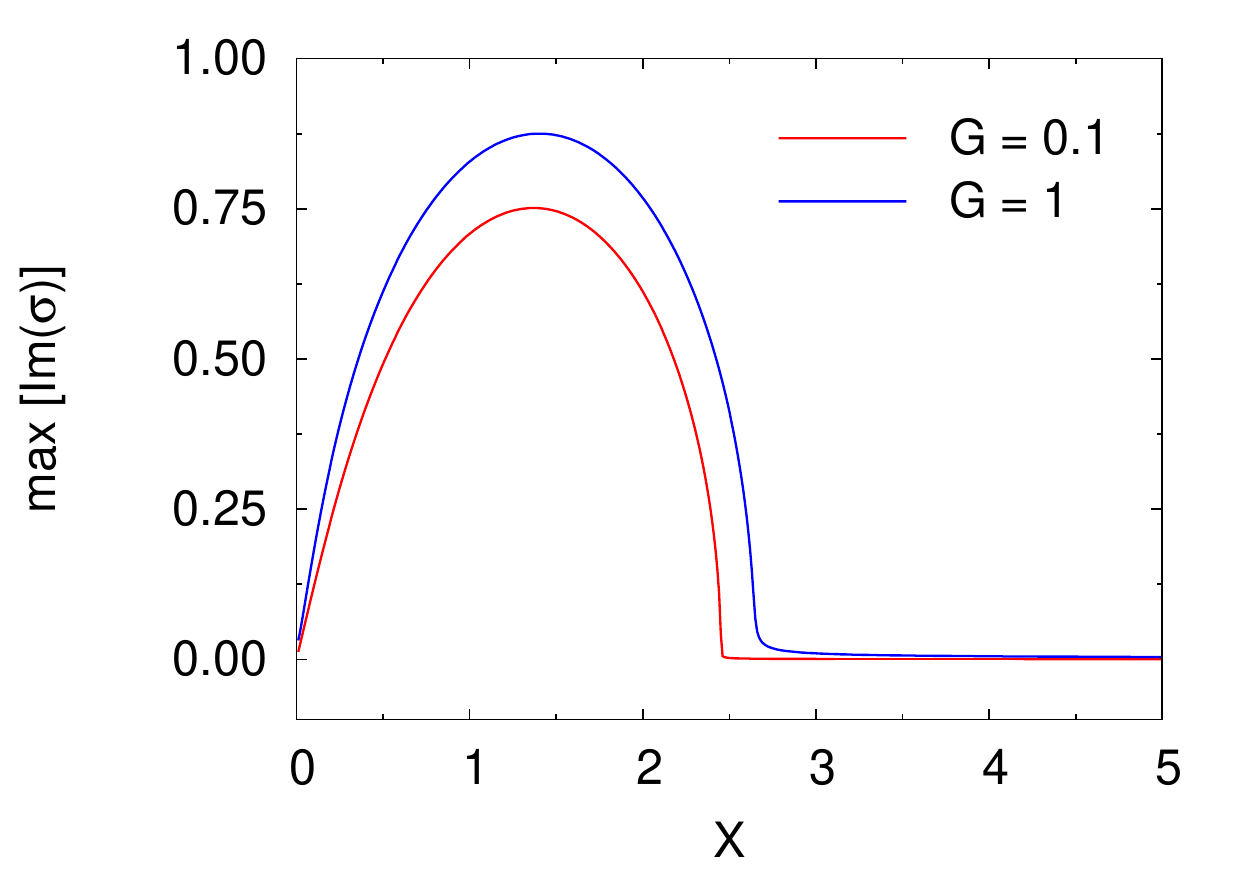}
  \end{minipage}
  
  \begin{minipage}[htbp!]{0.4\textwidth}
    \includegraphics[width=\linewidth]{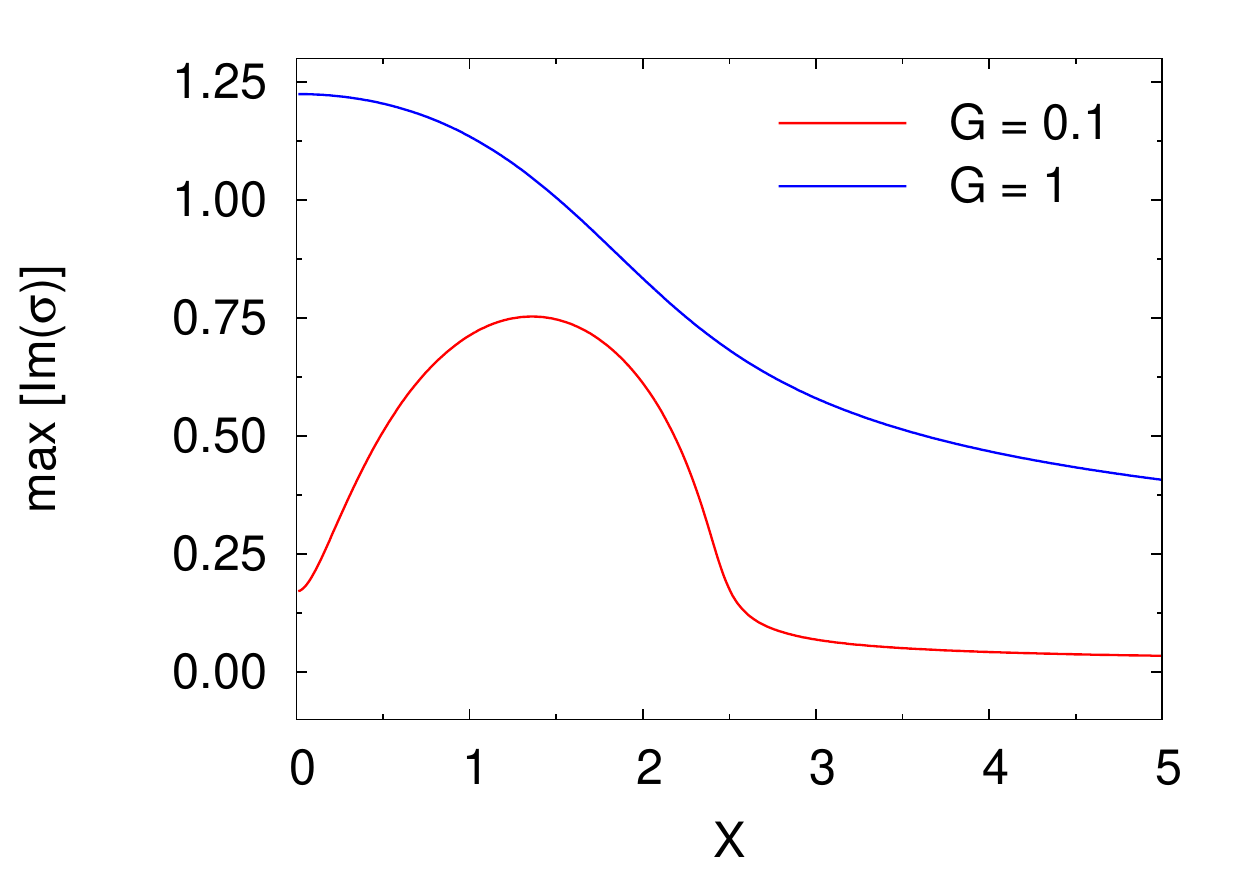}
  \end{minipage}
  \hspace{0.7cm}
 \begin{minipage}[htbp!]{0.4\textwidth}
    \includegraphics[width=\linewidth]{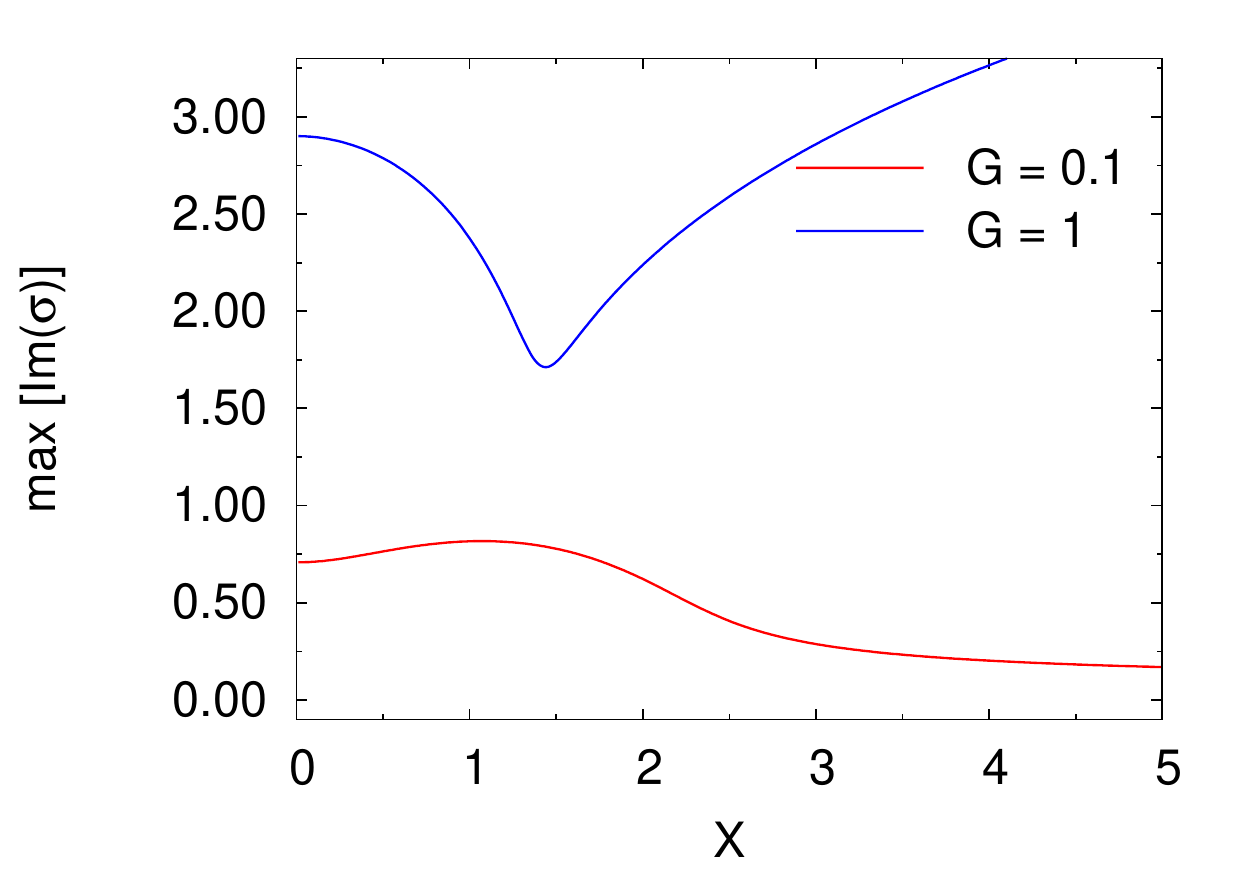}
  \end{minipage}
  \caption{The maximum growth rates attained for $m=0$ as a function of the dimensionless parameter $X$ for various values of the
    parameter $\kappa$ when the diamagnetization parameter $\epsilon$ is set to its maximum value of 0.5. 
    In each plot, $M_{A}=1$ and the radial gradient of the magnetic field, $G$, assumes two values: 0.1 (red lines) and 1 
(blue lines). The plots correspond to $\kappa$ values of 0 (top-left); 0.01 (top-right); 1 (bottom-left) and 5 (bottom-right). 
For $\kappa=0$ and $\kappa=0.01$ the growth rates are similar to those found for ideal MRI while for $\kappa>0.01$ 
the gradient in the magnetic field leads to higher growth rates.}
 \label{fig:Figure02}
\end{figure*}

In Fig.\ref{fig:Figure02} we show the maximum growth rates attained as a function of the dimensionless parameter $X$
for various values of the parameter $\kappa$ when the diamagnetization parameter $\epsilon$ is set to its maximum
value of 0.5. In each plot, $M_{A}=1$ 
and the radial gradient of the magnetic field, $G$, assumes two values:
0.1 (red lines) and 1 (blue lines). We see that when $\kappa=0$, i.e. $\bsr=0$, the maximum growth rates 
for the instability are only slightly larger than those of the ideal MRI (BH91). The effect of 
increasing $G$ values manifests itself in both the amplitude of the maximum growth rate and in the $X$ values for 
which the instability diminishes. For $\kappa=0.01$ the situation is almost indiscernible from the $\kappa=0$ case.
When $\kappa=1$ the maximum growth rate of the instability is much more sensitive to the gradient of the magnetic 
field, $G$. Although 
for $G=0.1$ the maximum growth rate is similar to that of the $\kappa=0$ case, for $G=1$, the maximum amplitude becomes 
larger than 1 and causes the instability to operate at larger wave numbers. This can be understood by considering the 
reshaping of the geometry of the magnetic field lines. In Figure 7 of DP07 we depicted the location of 
the fiducial radius $R_{0}$ wherein the various processes take place. The directions of the axis of rotation 
of the disc, $\Omega$; the vertical magnetic field, $B_z$; the gradient of the $B_z$ component of the magnetic 
field, $\nabla{B_z}$; electron drift velocity caused by the gradient and the curvature of the magnetic field, 
$v_{\nabla{B}+R_c}$, where $R_c$ is the unit vector of the radius of curvature. The curvature in the magnetic field lines is caused 
by the diamagnetic current. This current and the drift 
velocity of the frozen-in electrons bend the magnetic field lines and give  them a curvature as a result of which 
the $B_{R}$ component of the magnetic field is generated. Finally, for $\kappa=5$ we see that the maximum growth 
rate of the instability takes ever larger values, especially when $G=1$. For this value of the $\kappa$ and the range 
of $X$ values we consider, the instability does not vanish and the disc is unstable against the perturbations with 
all the wavenumbers.

\begin{figure*}
  \centering
\begin{minipage}[htbp!]{0.4\textwidth}
    \includegraphics[width=\linewidth]{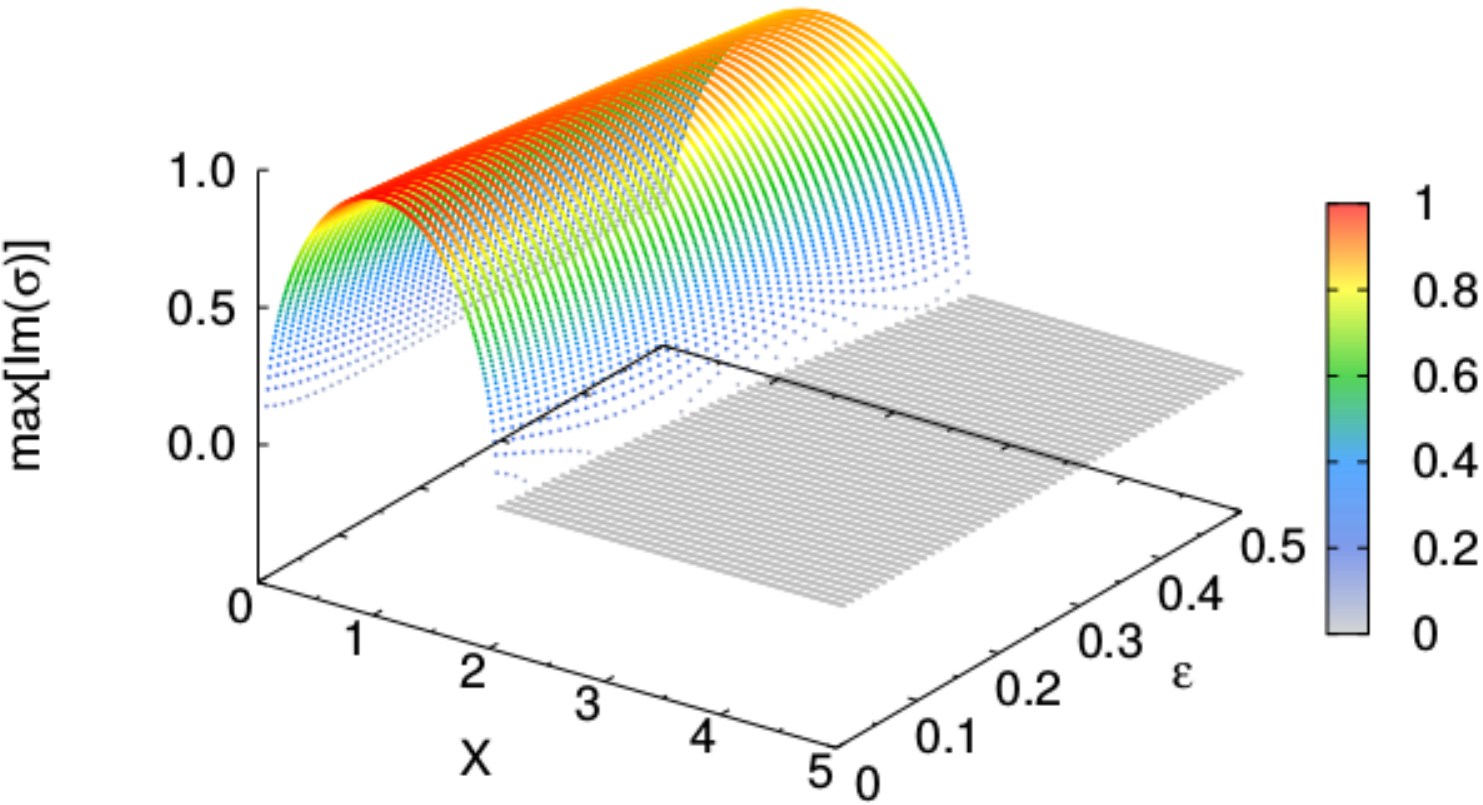}
  \end{minipage}
\hspace{1cm}
\begin{minipage}[htbp!]{0.4\textwidth}
  \includegraphics[width=\linewidth]{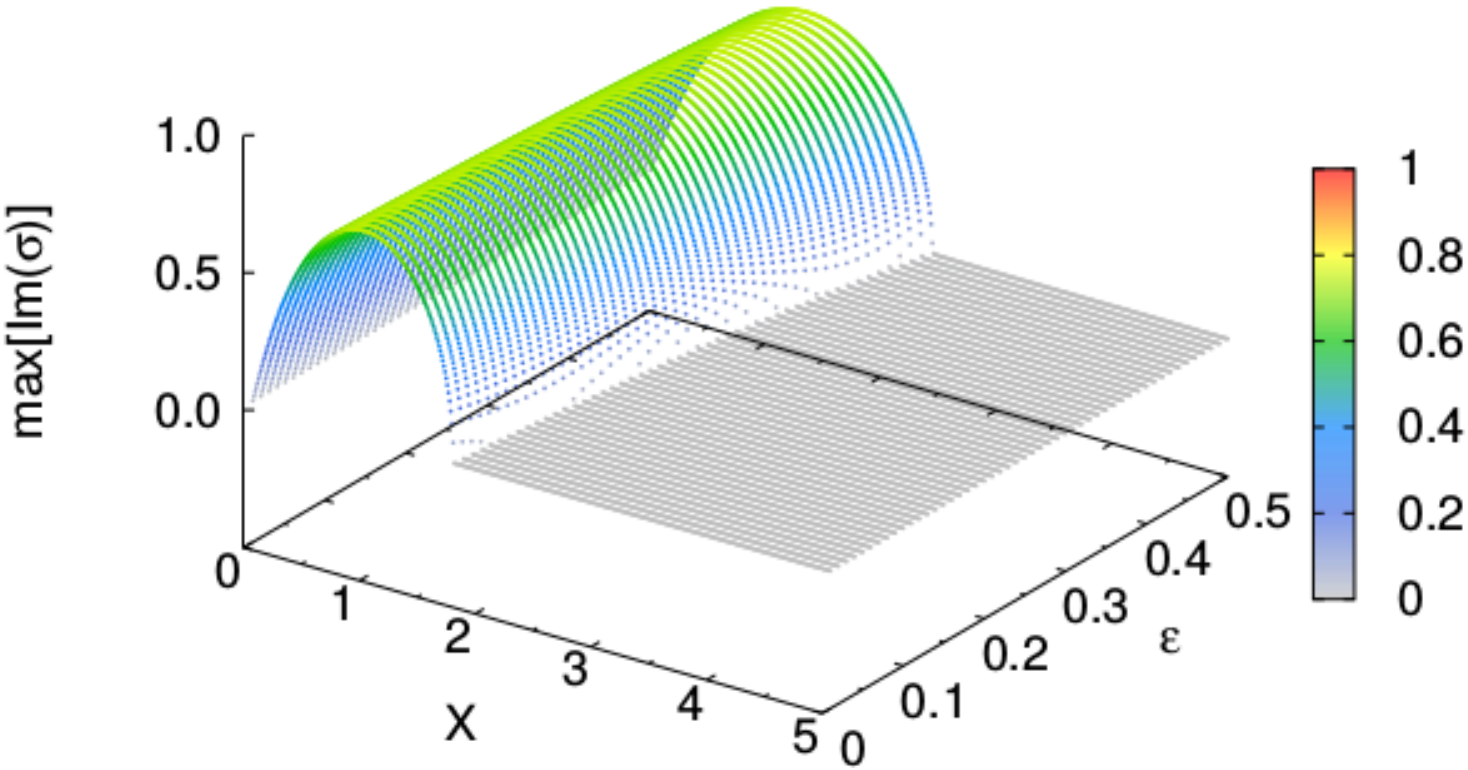}
\end{minipage}

 \begin{minipage}[htbp!]{0.4\textwidth}
    \includegraphics[width=\linewidth]{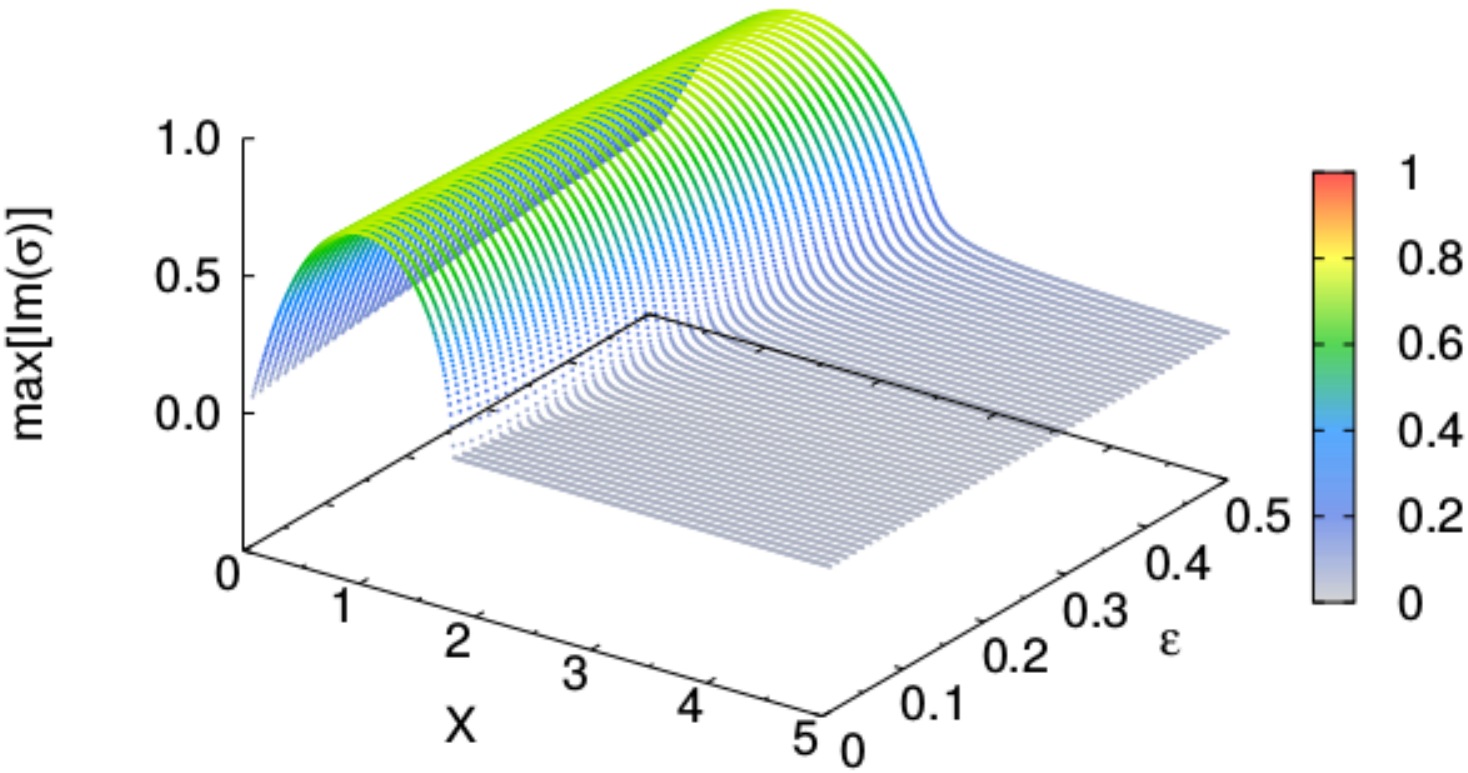}
 \end{minipage}
 \hspace{1cm}
  \begin{minipage}[htbp!]{0.4\textwidth}
    \includegraphics[width=\linewidth]{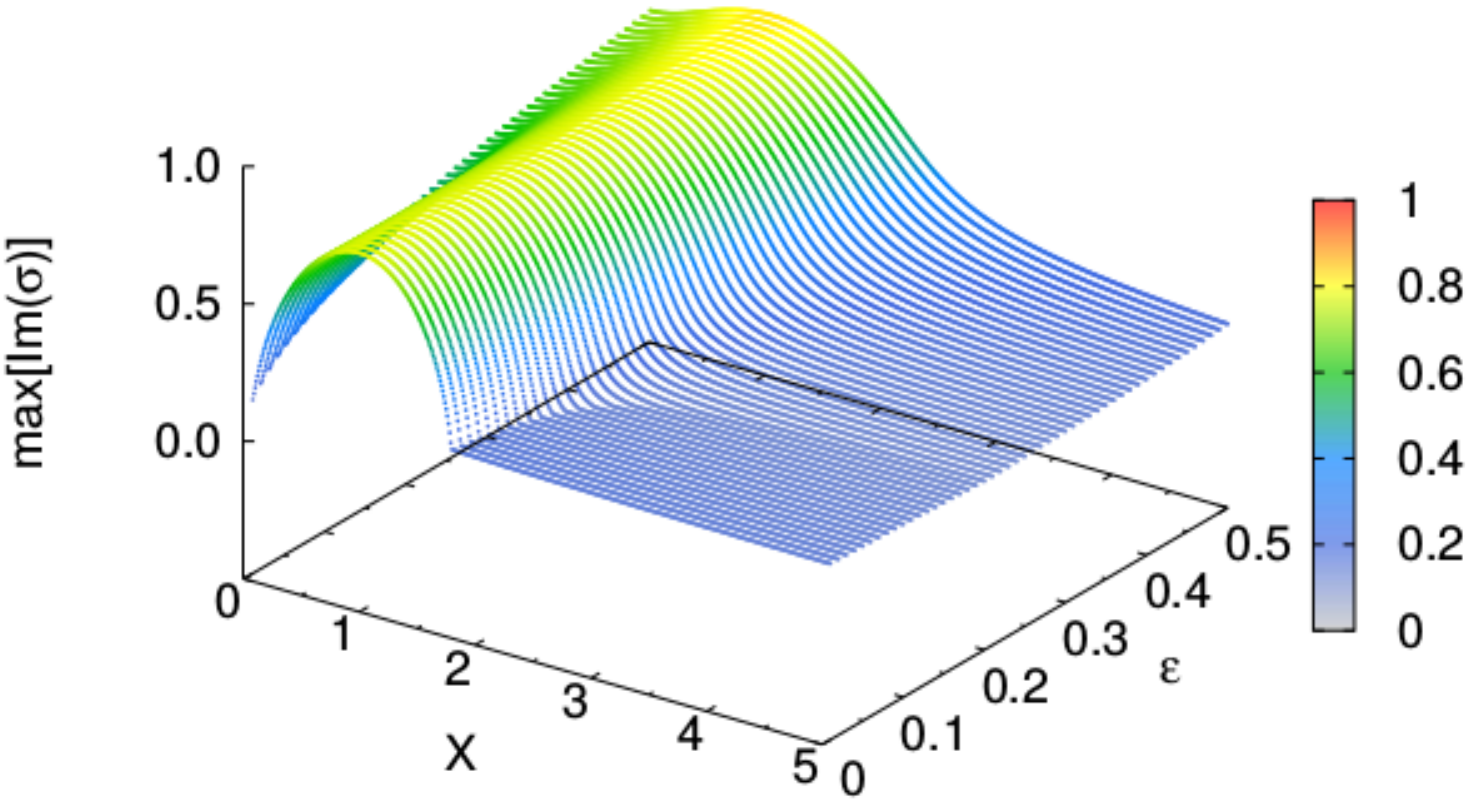}
  \end{minipage}
 \caption{ The maximum growth rates attained for $m=0$ for varying $\epsilon$ and $X$ when $G=0.1$.  Top-left: $\kappa = 0$, 
top-right: $\kappa = 0.01$, 
   bottom-left: $\kappa = 1$, bottom-right: $\kappa = 5$. The color-code indicates the amplitude of the growth rate. 
The maximum growth rate is achieved when $\kappa=0$. For $\kappa=5$ the unstable region shifts towards larger wavenumbers.}
 \label{fig:Figure03}
\end{figure*}

In Fig. \ref{fig:Figure03}  we show the maximum growth rates attained for varying $\epsilon$ and $X$
when $M_A=1$ and $G = 0.1$ for different values of the parameter $\kappa$. We see that when $\kappa=0$, the
value of the maximum growth rate increases towards a value of 1 as the magnetization
parameter $\epsilon$ increases. The growth rate for $\epsilon=0.5$ in this plot corresponds
to the one depicted by the red line ($G = 0.1$) shown in the top-left panel of Fig. \ref{fig:Figure02}.
When $ \kappa=0.01$, i.e. the radial component of the magnetic field is much smaller than the vertical 
one, the maximum value
of the growth rate is almost independent of the magnetization  parameter $\epsilon$, however the instability 
extends towards higher wavenumbers.
This can be inspected by the increase of the radius of the ridge-like
distribution of the growth rates towards higher $\epsilon$. When the radial and the vertical components of the magnetic field
are of the same magnitude, i.e. $\kappa=1$,  we see that although the instability diminishes sharply for
small $\epsilon$, it starts to diminish more smoothly at higher wavenumbers for higher $\epsilon$ values.
For $\kappa=5$ the most striking result is that the disc is unstable for all the wavenumbers we consider.
This can be seen by the colour-code in the growth rate.

\subsection{Non-axisymmetric mode}
\label{sec:Nonaxisymetric}
In the presence of shear, the non-axisymmetric perturbations are complicated. The radial wavenumber 
may increase linearly with time. In this case the effect of non-axisymmetric perturbations may be examined 
using shearing sheet approximation where the radial wavenumber is expressed as \citep{Goldreich65, Balbus92b}
\begin{equation}
    k_R(t)= k_R(0)-mt \frac{ d\Omega}{dR}.
\end{equation}
In the limit $k_R R >> m$, the time scale for the amplification of the perturbation is much  longer than 
the orbital time,  therefore the radial wavenumber, $k_R$, may be considered to be independent of time \citep{Kim00}. 
This condition is also satisfied when $|{\bf{k}}| \longrightarrow  \infty$ \citep{Terquem96}. For simplicity, in our 
analysis we do not consider perturbations in the radial direction, i.e. the perturbations are of the form 
${\rm {exp}}({i{\bf{k}}\cdot{\bf{r}}- i \omega t})$,
where ${\bf{k}}=(0,m/R,k_{z})$.

We investigate non-axisymmetric perturbations focusing particularly on the $m=1$ mode.
The coefficients in the dispersion relation are fairly complicated functions of the parameters involved in 
our problem. We therefore give the full expression in Appendix \ref{sec:fullDR}.

In Fig. \ref{fig:Figure04} we show the maximum growth rates for $m=1$ when the other set of parameters are 
identical to those in Fig. \ref{fig:Figure02}. 
When $\kappa=0$ the maximum  growth rate drops from $\sim 0.75$ (for $m=0$) to $\sim 0.5$
for $G = 0.1$. When $G =1$, the maximum growth rate has a higher value than for $G = 0.1$. The growth rate exhibits 
a plateau-like behaviour up to $X \sim 1$ which then decreases and vanishes at larger wavenumbers compared to the $m=0$ case.
For $\kappa=0.01$ the growth rates behave similar to those for the $\kappa=0$ case although the instability 
diminishes slightly more softly. 
For $\kappa=1$ and $\kappa=5$, all the waves are unstable in the $X$ range shown here. We note that we chose 
to depict the results of these models also up to $X=5$ in order to make the comparison with the $m=0$ modes 
easier. However, we have calculated the growth rates up to much higher $X$ values and saw that the instability 
diminishes beyond $X \sim 50$ and $X \sim 100$ for 
$\kappa=1$ and $\kappa=5$, respectively, for $G=0.1$. However, for $G=1$, the growth rates do not drop to 
zero for even large X values (not shown here).
\noindent
\begin{figure*}
\centering
  \begin{minipage}[htbp!]{0.4\textwidth}
    \includegraphics[width=\linewidth]{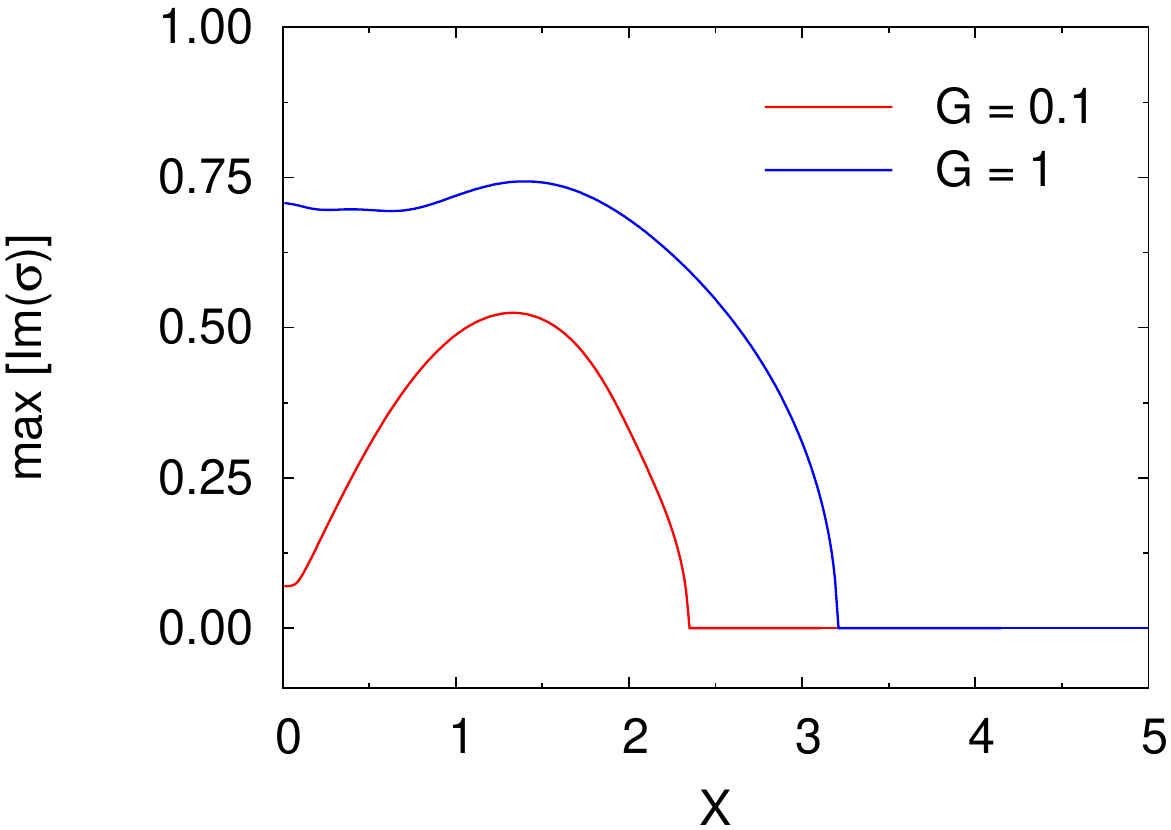}
  \end{minipage}
  \hspace{0.7cm}
  \begin{minipage}[htbp!]{0.4\textwidth}
    \includegraphics[width=\linewidth]{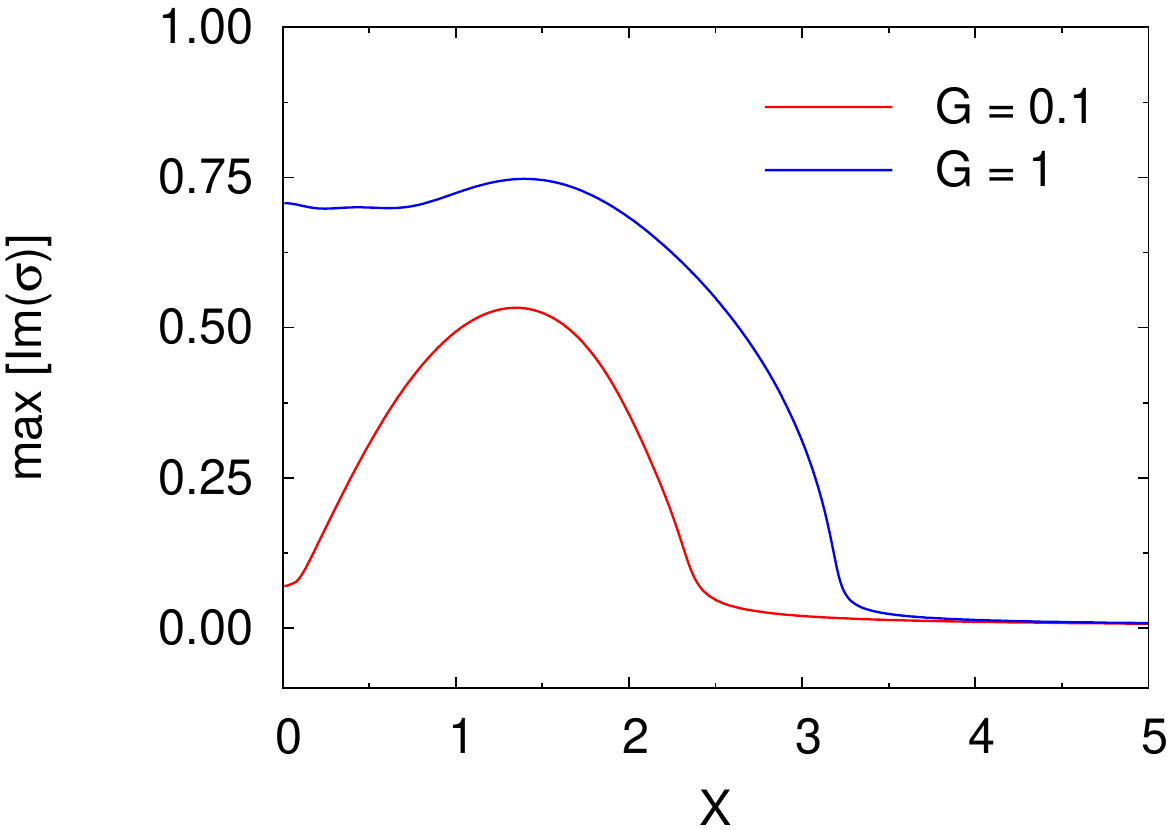}
  \end{minipage}

  \begin{minipage}[htbp!]{0.4\textwidth}
    \includegraphics[width=\linewidth]{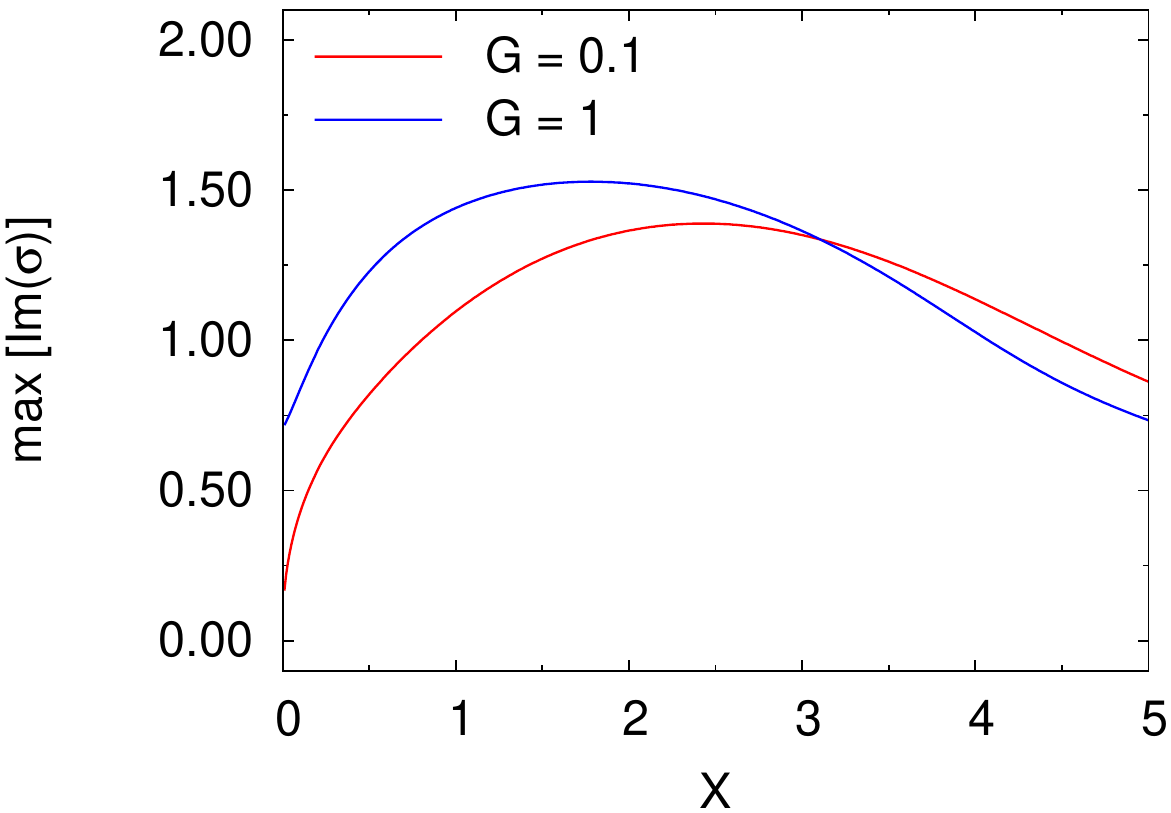}
  \end{minipage}
  \hspace{0.7cm}
  \begin{minipage}[htbp!]{0.4\textwidth}
    \includegraphics[width=\linewidth]{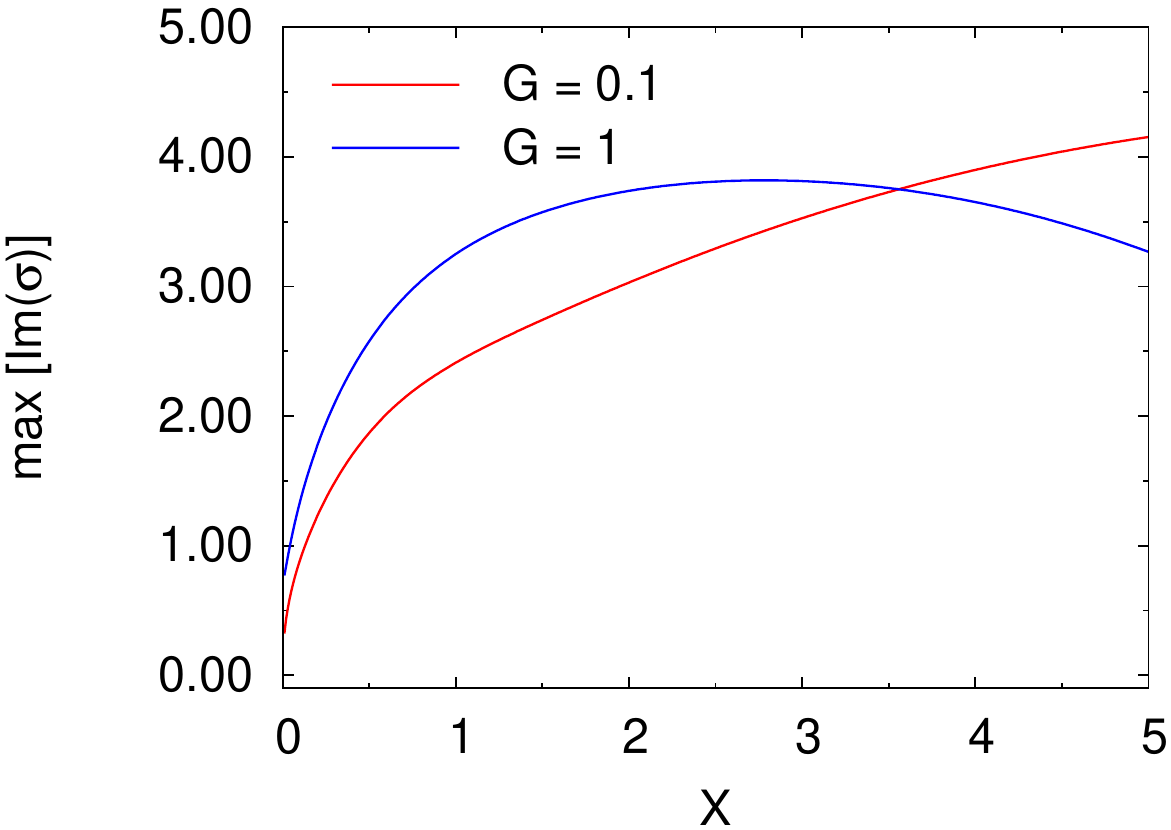}
  \end{minipage}
  \caption{As in Fig. \ref{fig:Figure02} but for a wave mode of $m=1$. It can be seen that the maximum
    growth rates are smaller than those of the $m=0$ mode for $\kappa=0$ and $\kappa=0.01$ when $G=0.1$. 
For both $\kappa=1$ and $\kappa=5$ the growth rates 
are much larger than those for the $m=0$ mode and all the waves are unstable in the X range considered.}
 \label{fig:Figure04}
\end{figure*}

\begin{figure*}
\centering
  \begin{minipage}[htbp!]{0.4\textwidth}
    \includegraphics[width=\linewidth]{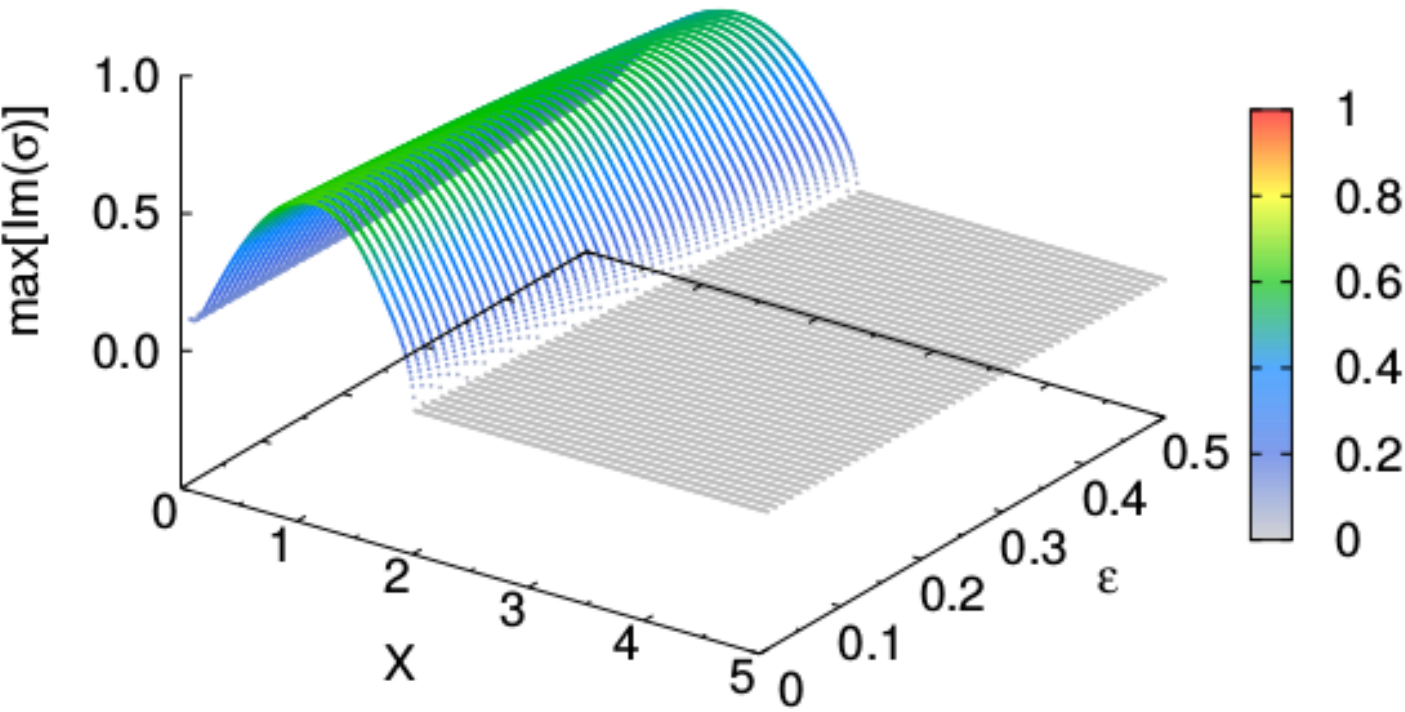}
  \end{minipage}
\hspace{1cm}
\begin{minipage}[htbp!]{0.4\textwidth}
  \includegraphics[width=\linewidth]{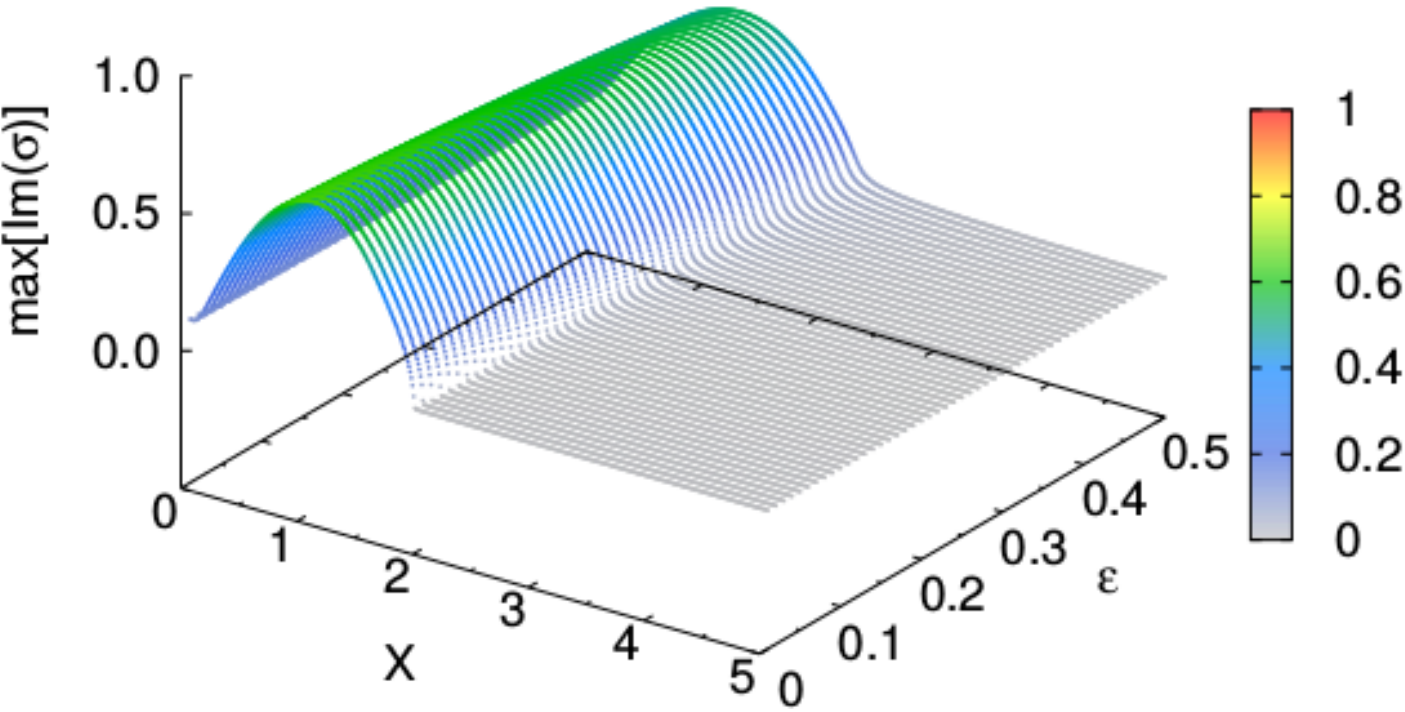}
\end{minipage}
  
 \begin{minipage}[htbp!]{0.4\textwidth}
    \includegraphics[width=\linewidth]{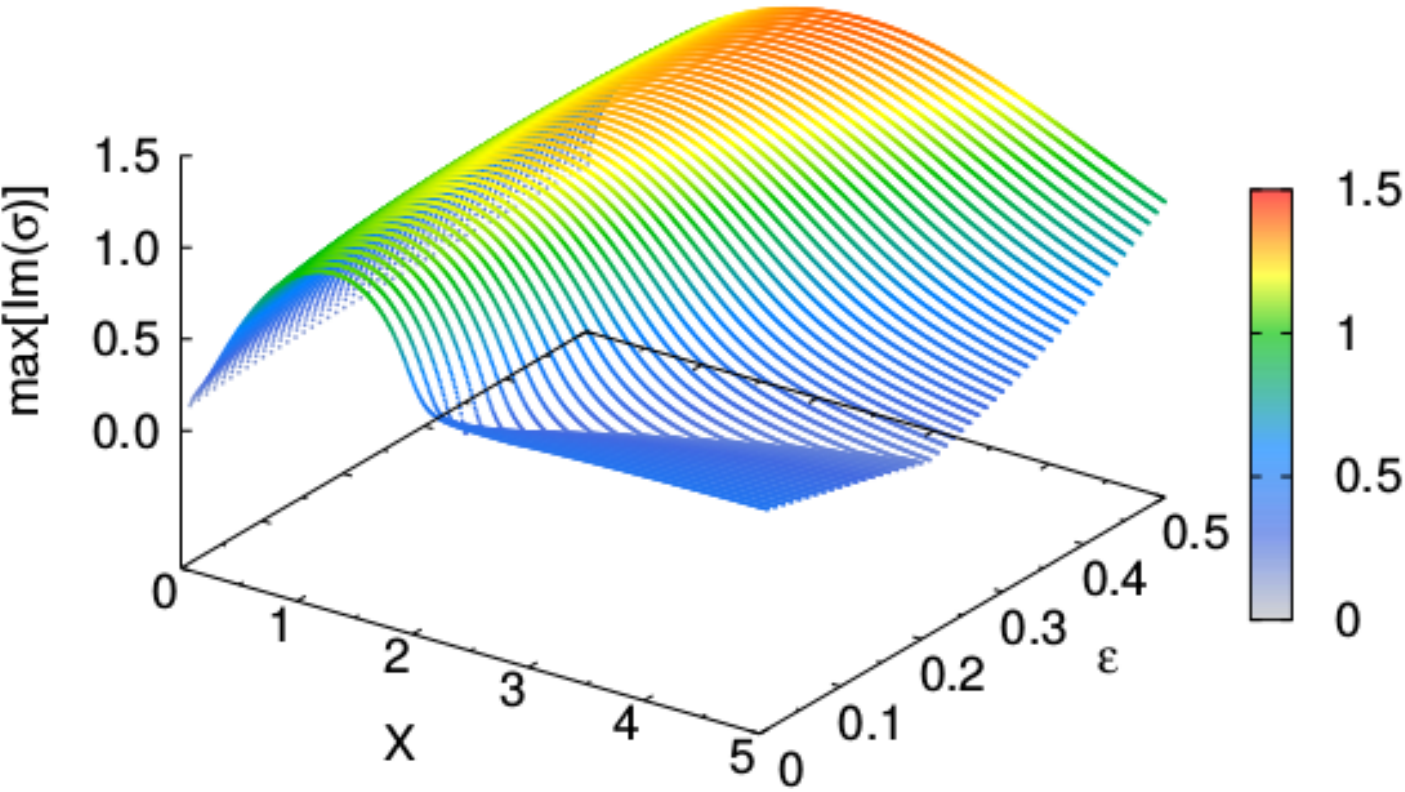}
 \end{minipage}
 \hspace{1cm}
  \begin{minipage}[htbp!]{0.4\textwidth}
    \includegraphics[width=\linewidth]{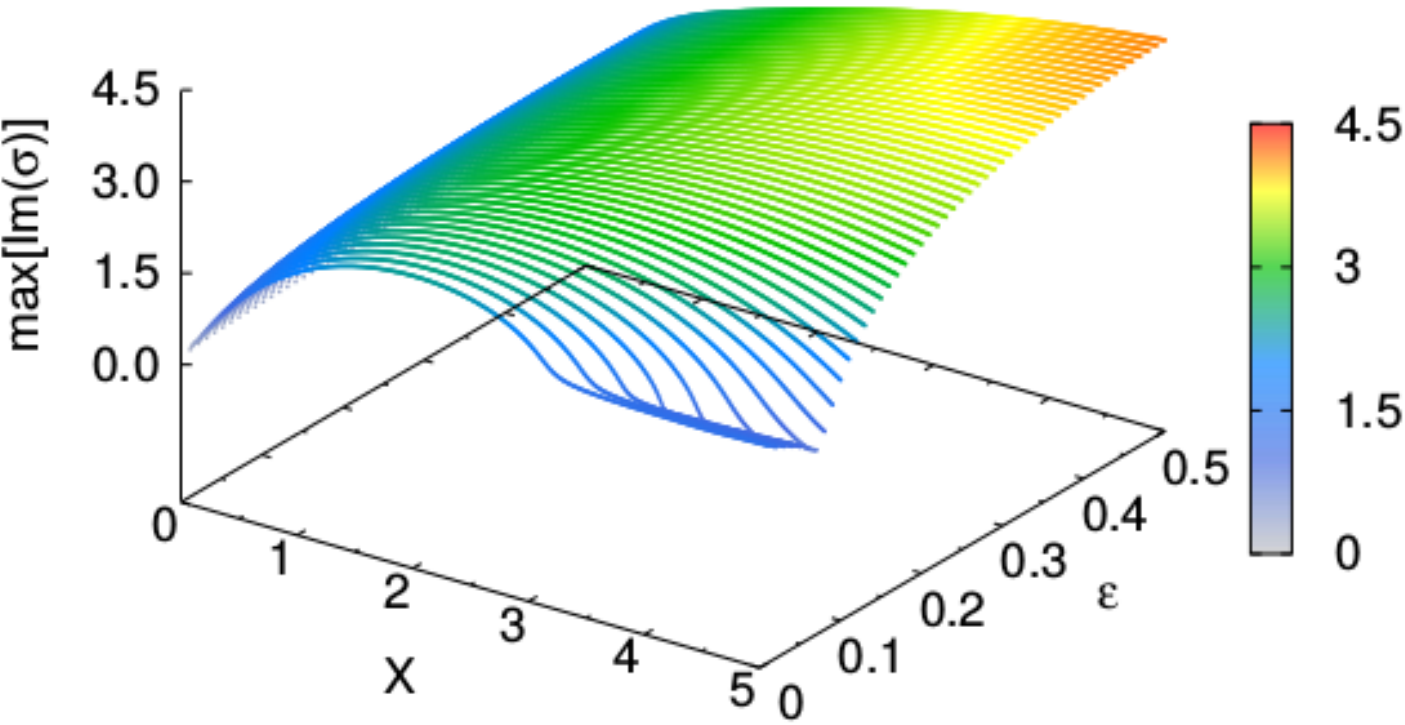}
  \end{minipage}
  \caption{As in Fig. \ref{fig:Figure03} but for a wave mode of $m=1$. The maximum growth rates decrease with
    increasing $\epsilon$ for $\kappa=0$ and 0.01. For $\kappa > 1 $ and large $\epsilon$ all the 
waves are unstable in the $X$ range shown.}
 \label{fig:Figure05}
\end{figure*}

Fig. \ref{fig:Figure05} depicts the maximum growth rates for varying $\epsilon$ and $X$ for the $m=1$ mode. The other 
parameters are identical to those in Fig. \ref{fig:Figure03}. When $\kappa=0$, contrary to the case for $m=0$ where 
the magnetization strengthens the instability, the maximum growth rates display a decreasing 
profile with $\epsilon$ when $m=1$. A similar behavior of the growth rate is observed for $\kappa=0.01$. The 
situation changes however when $\kappa > 1$. For these modes the growth rates assume ever increasing values for 
increasing $\epsilon$.  In our figures wherein $\kappa$  is assumed to have values 1 and 5 the growth rate of the 
unstable waves much exceeds the ideal MRI value of 0.75 for large $\epsilon$ values.

\section{Discussion}
\label{sec:discussion}

\subsection{The effects of non-ideal terms}
In our calculations, we neglected the effects of the non-ideal terms arising from Ohmic diffusion, the Hall effect and ambipolar diffusion on the growth rates of the MRI. Roughly,
Ohmic diffusion acts at the inner regions of protoplanetary discs, the Hall effect at intermediate 
radii and the ambipolar diffusion operates at larger distances from the central star and at the surface layers 
of the disc (\ca{Armitage11} \cy{Armitage11} and references therein). Since the diamagnetic effect is most pronounced 
in the inner regions of the disc, one might think that its interplay with the Hall effect and Ohmic diffusion might modify the growth rates to values other than reported here. We defer the inclusion of these two non-ideal terms in our model to a future work. The effects of these terms in the absence of the diamagnetic effect have been studied in detail by several authors.

The MRI in protostellar discs in the presence of Ohmic diffusion can operate when the growth rate of the instability is larger than the diffusion rate. The instability is suppressed 
when the magnetic Reynolds number $\rm Re_{\it{M}} < 1$ \citep{Jin96, sano99}.
\cite{Balbus01} searched the linear stability of a magnetized protoplanetary disc with the Hall effect present. They found that the Hall effect allows discs with increasing 
rotation profiles to become unstable as well. They also reported that if the magnetic field has a
radial component, one can always find certain wavenumbers leading to disc instability. 
\cite{Kunz13} investigated local, three dimensional, resistive Hall-MHD simulations of the 
MRI with a condition where Hall effect is dominant over Ohmic dissipation. Their linear stability 
analysis showed that the MRI grows exponentially at the regimes where the result of resistive 
MHD indicates stability. However, despite the exponential growth, the MRI saturates, forms the  
so-called {\it{zonal fields}} and {\it{zonal flows}} in the magnetic and velocity fields, 
respectively, as a result of which the transport of angular momentum diminishes. 

When the ion-neutral collision timescale is smaller than the Keplerian orbital 
time scale ${\Omega}^{-1}$, ambipolar diffusion may be the dominant non-ideal effect \citep{Armitage11}.
The combined effects of these non-ideal terms have also been studied in several simulations indicating
that when the magnetic field and the disc rotation are not aligned, the discs are more prone 
to MRI \citep{Bai11, Bai15}. In a detailed investigation of linear disturbances at weakly ionized, 
magnetized planar shear flows, \cite{Kunz08} showed that the combination of ambipolar diffusion, the Hall effect, 
and the shear triggers instability in accretion discs. Although Ohmic diffusion and ambipolar diffusion act 
to suppress the activity of MRI, Hall effect acts to activate MRI by producing an azimuthal magnetic field 
which prevents the formation of {\it{zonal fields}} \citep{Lesur14}. At distances $r \lesssim 1$ AU from the 
central star and at the outer regions where ambipolar diffusion becomes dominant, the orientation of the magnetic 
field with respect to the disc's rotation doesn't play a role in angular momentum transport, however between 
5 AU $< r < $ 10 AU where Hall effect is dominant,  discs with $B \cdot \Omega < 0$ may exhibit 
bursts of accretion. Depending upon the chemical model used in the simulations, presence or absence of 
bursts reveal the comprehensible regions of protoplanetary disc \citep{Simon15}.
We defer the inclusion of these non-ideal terms to our model of diamagnetic protostellar 
discs to a future work.

\subsection{Torques and disc warping}

Observations of protoplanetary discs indicate that their rotation axis might be
misaligned with respect to the magnetic field axis \citep{Li16} and that these discs might be warped or 
broken. These effects have consequences on the evolution of the discs and the spins of the central 
stars. The evidence for warping comes from the periodic dimming of the light curves of the central stars 
\citep{Bouvier07}, from kinematic modelling of molecular lines tracing the inner regions of the discs 
\citep{Rosenfeld12} or from modelling the scattered light images \citep{Benisty18, Facchini18}. 
The magnetic interaction between the central star and the disc may alter the spin direction of the 
central star towards misalignment with respect to the disc as indicated by several observations \citep{Lai11}. 
The time-scale by which the disc settles into a steady warped disc configuration is determined by the 
competition between the magnetic and the internal viscous torques \citep{Foucart11}.

One of the plausible mechanisms for disc warping is the magnetic interaction between 
the central star and the surrounding disc. This interaction inevitably brings about
a surface electric current on the disc and the shape of the current depends on various factors like the 
presence of the magnetic field perpendicular to the disc, diamagnetic property of the disc and the dissipative 
processes in the magnetosphere \citep{Lai03}. The surface current on the disc and the horizontal magnetic field 
of the stellar dipole cause disc warping and precessional torques (\ca{Lai99} (\cy{Lai99}; \cy{Lai03})). 
The force generating the torque on the disc is given as ${\bf{F}}_{z}= {\bf{J}}_{\phi}\times{\bf{B}}_{R}$, 
that is, the diamagnetic current flowing in the azimuthal direction interacts with the radial component 
of the stellar dipole magnetic field and the resulting force generates the torque. 
\cite{Paris18} studied the dynamics of warped magnetized discs taking into 
account the vertical component of the magnetic field. They found that even if the disc is stable against 
MRI for large scale magnetic fields, its dynamics might still be altered by the warping of the disc.

In our calculations, we considered the interaction between a magnetic star and a surrounding disc to 
obtain the MRI growth rates taking into account the radial component of the magnetic field, $B_{R}$. 
The greater the bending of the field lines, in other words, the smaller the radius of curvature the 
bigger the magnitude of the radial component becomes. As a result, the magnitude of the force
${\bf{F}}_{z}$, generating torque on the disc may come closer to the point when the warp of the 
disc is generated. 

\section{Summary and conclusions}
\label{sec:summary}
We considered the interaction between a magnetic star and a surrounding diamagnetic protostellar disc 
which is misaligned with respect to the magnetic moment axis. We carried out a linear stability analysis and 
obtained the growth rates of the MRI for different values of the magnetic field strength, the diamagnetization 
parameter and the strength of the radial gradient of the magnetic field for both axisymmetric ($m=0$) and 
non-axisymmetric ($m=1$) perturbations. Our findings can be summarized as follows:

\noindent
($i$) For the axisymmetric case when $\kappa =0$, the instability criterion differs from MRI one. The magnetic 
field gradient resulting from the diamagnetic current causes the disk become more unstable. Also this gradient  
increases the growth rate of the instability.

\noindent
($ii$) For the $m=0$ mode, the maximum growth rate of the MRI has a value similar to the 
ideal MRI value of 0.75 when the radial gradient of the vertical component of the magnetic field is 
small and the parameter $\kappa < 1$. A steeper change in the gradient of the magnetic field leads to a 
slight increase in the growth rates. For $\kappa > 1$, the growth rates assume ever increasing values and all 
the  waves are unstable.

\noindent
($iii$) The maximum growth rates for the $m=1$ mode are smaller than those of the $m=0$ mode, especially 
for a small value of $G$ and $\kappa < 1$.

\noindent
($iv$) For maximum magnetization, $m=1$ and $\kappa > 1$ all the waves are unstable for a large magnetic field gradient.

\section*{Acknowledgments}
A.U. thanks the faculty of the Department of Astronomy and Space Sciences at Ege University for their hospitality 
where part of this work has been completed. 

\bibliographystyle{mnras}
\bibliography{MRI}

\onecolumn
\appendix
\section{Terms of the full dispersion relation}
\label{sec:fullDR}
Defining  $y = m / (Rk_z)$ we write the general dispersion relation for a non-axisymmetric 
perturbation with $m \ne 0$ as
\begin{equation}
{\sigma}^5 + a {\sigma}^4 + \frac{1}{y^2+1} \Big\{ b {\sigma}^3 + c {\sigma}^2 + 
d {\sigma} + e \Big\}= 0.
\label{eq:fullDR}
\end{equation}
Here

\begin{equation}
a = -5 m,
\label{eq:fullDRA}
\end{equation}

\begin{align}
\begin{split}\label{eq:fullDRB}
b = {}&-\frac{G^2 (\epsilon-1)(y^2+1)}{{M_A}^2} +i \frac{\kappa G X (\epsilon -1 -y^2 (1+3 \epsilon))}{M_A}-2X^2 (1-\epsilon)
-X^2 y^2 (1+\epsilon^2)+2X^2 y^2 \kappa^2 \epsilon (\epsilon-1) +10 m^2 (y^2+1) -\eta^2,
\end{split}\\
\begin{split}\label{eq:fullDRC}
c  ={} & -\frac{2 G^2 y (-\frac {3}{2}y m + i \kappa ) (\epsilon-1)} {{M_A}^2}  - \frac{4 X G (\epsilon-1)^2 y \kappa^2}{M_A} + 
         \Big( i \frac{3 X G (1+3 \epsilon) m y^2 }{M_A} - i \frac{X ((\epsilon-1) (4+3G)+\eta^2 \epsilon)m}{M_A}  \Big) \kappa  \\
       & + \frac{1}{2} \frac{X G (\epsilon-1) (6 G - \eta^2 - 4 \epsilon -4)y}{M_A} + 6 X^2 \epsilon (\epsilon-1) m y^2 \kappa^2 + 
         (3 X^2 (1+\epsilon^2) m-10 m^3) y^2 -10 m^3 + (6 X^2 (1-\epsilon)+3 \eta^2)m,
\end{split}\\
\begin{split}\label{eq:fullDRD}
d = {}& i \frac{\kappa G^3 X (y^2 (\epsilon^2-1)-2 \epsilon (\epsilon-1))}{{M_A}^3} 
+ \Big( \frac{2 G^2 X^2 (1+\epsilon^2 - 4 \epsilon) y^2 }{{M_A}^2} + \frac{2 G^2 \epsilon
  (2 \alpha + X^2 (\epsilon+1))}{{M_A}^2}  \Big) \kappa^2 \\
      & +  \Big( \frac{G^2 X^2 (\epsilon-1)}{{M_A}^2} - \frac{3 G (\epsilon-1)m^2}{{M_A}^2}  \Big)y^2 
        - \frac{G^2 X^2 (\epsilon-1)^2}{{M_A}^2} - i \frac{2 G \epsilon X^3 (\epsilon-1) (\epsilon-3) y^2 \kappa^3}{M_A} \\
        & + \Big[ \Big(  -i \frac{3 X G (1+ 3 \epsilon) m^2}{M_A} + i \frac{4 G \Big( -\frac{1}{4} (\epsilon-2)
            (\epsilon^2 + 4 \epsilon -1) \Big) X}{M_A} \Big) y^2  
        + i \frac{G ((\epsilon-1)^2 X^2 + 2 \alpha (1-3 \epsilon))X}{M_A} \Big] \kappa \\
      & + (-6 X^2 \epsilon (\epsilon-1) m^2 +8 X^2 G (\epsilon-1)^2 - 2 X^4 \epsilon (\epsilon-1) (\epsilon-3))y^2 \kappa^2 + i X^2 ((\epsilon-1)(3 G + 8)+2 
        \eta^2 \epsilon)m y \kappa \\
      & + (5 m^4 - 3 X^2 (1+\epsilon^2) m^2 + 4 X^2 G (\epsilon^2 -1) + 3 G^2 X^2 (1-\epsilon) + 
        X^4 \epsilon (1-\epsilon^2)+X^2 G \eta^2 (\epsilon-1))y^2 + (6X^2 (\epsilon-1)-3 \eta^2)m^2 \\
      & + 5 m^4 + 2 X^2 \alpha (1-\epsilon)+X^4 (1-\epsilon)^2, 
\end{split}\\
\begin{split}\label{eq:fullDRE}
e = {}& -\frac{\kappa G^3 y X (- i m y (1-\epsilon^2) + 2 \kappa \alpha \epsilon (1-\epsilon))}{{M_A}^3}
        - \frac {2 G^2 X^2 (1+\epsilon^2 - 4 \epsilon) m y^2 \kappa^2 }{{M_A}^2} 
        + i \frac {2 \Big ( \Big ( G-\frac{3}{2} \alpha \Big ) \epsilon + \frac{1}{2} \alpha \Big ) (\epsilon-1) G^2 X^2 y \kappa}{{M_A}^2}\\
      & - \frac{(\epsilon-1) m G^2 (-m^2 +X^2 (\epsilon+1))y^2 }{{M_A}^2} + i \frac{4 \kappa^3 G^2 \alpha \epsilon^2 y X^2}{{M_A}^2} 
        + i \frac{2 G X^3 \epsilon (\epsilon^2 - 4 \epsilon +3) m y^2 \kappa^3}{M_A} \\
      & - \frac{2 G X ((\alpha (1-3 \epsilon)+G(\epsilon+1))X^2+ 2 \alpha G)y \kappa^2}{M_A}  + \Big [ \Big( -\frac{i G X (2 G (\epsilon-1) 
        - X^2 (\epsilon-2)(\epsilon^2 + 4 \epsilon-1) )m}{M_A} + \frac{i G X (1+ 3 \epsilon)m^3}{M_A} \Big)y^2  \\
      & - \frac {i X (2 \alpha \epsilon X^2 (\epsilon-1) - 2 \alpha G (3 \epsilon-1) + X^2 G (\epsilon-1)^2 )m }{M_A} \Big ]\kappa  
        + \frac{G X^3 (\epsilon-1)^2 (G-\alpha)y}{M_A} \\
      & +(2 X^2 \epsilon (\epsilon-1) m^3 + (2 X^4 \epsilon (\epsilon-1) (\epsilon-3) 
        - 4 X^2 G (\epsilon-1)^2 )m)y^2 \kappa^2 \\
      & + \Big( X^2 (1+\epsilon^2) m^3 -m^5 +\frac{1}{2} X^2 G \eta^2 (1-\epsilon) 
        - G^2 X^2 (1-\epsilon)- X^4 \epsilon (1-\epsilon^2) + 2 X^2 G (1-\epsilon^2) \Big)m)y^2  \\
      & +(2 X^2 (1-\epsilon)+\eta^2)m^3 - m^5 + (-X^4 (1-\epsilon^2)-2 X^2 \alpha (1-\epsilon))m + i X^2 ((1-\epsilon)(G+4) - 
\eta^2 \epsilon) m y \kappa.
\end{split}
\end{align}

For the axisymmetric mode where $m=0$ (and therefore $y=0$) the coefficients $a$, $c$ and $e$  given in 
equations \ref{eq:fullDRA}, \ref{eq:fullDRC} and 
\ref{eq:fullDRE}, respectively, vanish and the dispersion relation reduces to the quartic given in
equation \ref{eq:dispersionm0}.

\label{lastpage}

\end{document}